\documentstyle[11pt,newpasp,twoside]{article}
\def\edcomment#1{\iffalse\marginpar{\raggedright\sl#1\/}\else\relax\fi}
\reversemarginpar

\begin{document}
\title{Feeding the active nucleus of NGC\,1097}
\author{Thaisa Storchi-Bergmann \& Rodrigo Nemmen da Silva}
\affil{Instituto de F\'\i sica, UFRGS, Porto Alegre, Brasil}
%\email{thaisa@if.ufrgs.br}

\author{Michael Eracleous}
\affil{Pensylvania State University,
University Park, PA 16802}

\begin{abstract}

We discuss the signatures of an evolving accretion disk
around the nucleus of NGC\,1097, based on observations
of its double-peaked H$\alpha$ emission-line profile which
cover a time span of 10 yrs. The
monitoring of the profile has allowed us
to better constrain the inner radius of the accretion
disk, and to propose a scenario in which the
ionizing source is getting dimmer and making the peak of
the disk emissivity to drift inwards.

\end{abstract}

Now that it is generally believed that supermassive black holes
(hereafter SMBH) may be present in most galactic bulges, an important
problem of AGN (active galactic nuclei) research is to
explain how the SMBH is fed. This feeding requires
in many cases the presence of an accretion disk,
whose signatures, are, nevertheless,
so far scarce. We discuss here the case of NGC\,1097 which,
similarly to $\sim$10\% of radio galaxies (Eracleous \& Halpern 1994),
shows a double-peaked H$\alpha$ emission-line profile whose
width is about twice the typical
for the broad line region of active galactic nuclei (AGN),
arguing for a different origin of these profiles.
The double-peaked profile of NGC\,1097 was first discovered in
1991 (Storchi-Bergmann et al. 1993), was then shown to be variable,
with a pattern which could be reproduced by a precessing elliptical 
accretion ring model with pericenter radii between 1300 and 1600 gravitational
radia R$_g$ (Storchi-Bergmann et al. 1995, 1997).

The case for an accretion disk has been recently strengthened by the
observed variability pattern over ten years of observations
of the nuclear H$\alpha$ profile.
In summary, we have observed that, besides showing the
previously known variation between the relative fluxes of blue and 
red peaks, these peaks are drifting apart, at the same time as the
total flux of the line is decreasing. In order to reproduce the most
recent data we had to decrease the inner radius of the ring
in the model from 1300 to 450\,R$_g$
(and thus more properly the ring can now be called a disk),
suggesting {\it  we were witnessing the
material from the disk falling towards the SMBH!} Nevertheless we
soon noticed that the time scale spanned by the observations was too
short by two orders of magnitude from the viscous time scale
of an accretion disk around the SMBH in  NGC\,1097 (estimated
to have a mass of 10$^6$M$_\odot$).

An alternative scenario is based on the fact that, at the same time
as the two peaks are drifting apart, the total flux in the line is decreasing.
This association suggests that the ionizing source is
getting dimmer, thus ionizing
regions successively closer to the nucleus and thus with higher
velocities in the accretion disk, making the two peaks to drift apart.
We could successfully reproduce the observations by a precessing
accretion disk model with a varying emissivity, such that the
radius of maximum emissivity $\xi_q$ is decreasing with time. Fig. 1 shows 
the first and last observed H$\alpha$ profile together with a corresponding
cartoon showing the orientation $\Phi_0$ and emissivity of the disk
in these two epochs.

Our monitoring and consistent
modeling show that the double-peaked H$\alpha$
emission in NGC\,1097 presents
many ingredients predicted in accretion models, including
the recent emergency of a wind,
%detected in the form of a blueshift of the whole line by
%$\sim-$500\,km\,s$^{-1}$,
strengthening the case of an accretion
disk as origin of the double-peaked profile.
A detailed analysis of the
observations and modeling of the disk
will be presented elsewhere.

\begin{figure*}
\vspace{10.0cm}
\caption{Observed profiles and fitted models (left)
together with corresponding cartoons (right) showing the change in
emissivity and orientation $\Phi_0$ (observer to the right)
of the disk between two epochs.}
%\label{var1}
\includegraphics{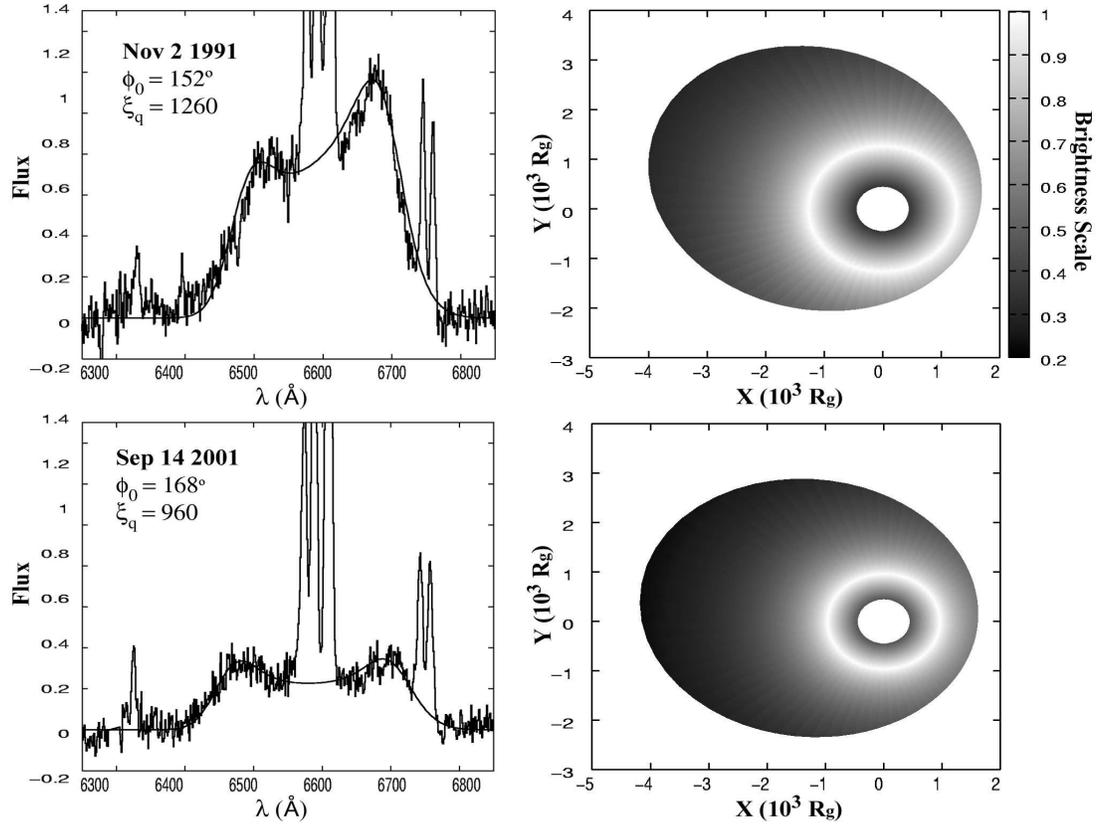}
%\special{psfile=esp_ngc6684.ps hoffset=170 voffset=30 hscale=35 vscale=35}
%\special{psfile=var_mrk34.ps hoffset=0 voffset=200 hscale=40 vscale=40}
%\special{psfile=var_mrk273.ps hoffset=260 voffset=200 hscale=40 vscale=40}
%\special{psfile=var_mrk348.ps hoffset=0 voffset=-10 hscale=40 vscale=40}
%\special{psfile=var_mrk477pa44.ps hoffset=260 voffset=-10 hscale=40 vscale=40}
\end{figure*}

\end{document}